\pdfoutput=1

\documentclass[twocolumn]{autart}    

\usepackage{graphicx}          
\usepackage{cite}
\usepackage{bm}
\usepackage{amsmath,amssymb}
\usepackage{mathrsfs}
\usepackage{amsfonts}
\usepackage{color}
\begin{document}

\begin{frontmatter}

\title{Expectation Synchronization Synthesis in Non-Markovian Open Quantum Systems\thanksref{footnoteinfo}} 

\thanks[footnoteinfo]{This paper was not presented at any IFAC 
meeting. Corresponding authors: Feng Pan, Kun Liu. }

\author[Beijing]{Shikun Zhang}\ead{zhang.shikun@outlook.com},    
\author[Beijing]{Kun Liu}\ead{kunliubit@bit.edu.cn},               
\author[Canberra]{Daoyi Dong}\ead{daoyidong@gmail.com},  
\author[Beijing]{Xiaoxue Feng}\ead{fengxiaoxue@bit.edu.cn},
\author[Beijing]{Feng Pan}\ead{andropanfeng@126.com}

\address[Beijing]{School of Automation, Beijing Institute of Technology, Beijing 100081, China}  
\address[Canberra]{ School of Engineering and Information Technology, University of New South Wales, Caberra ACT 2600, Australia}             


\begin{abstract}                          
In this article, we investigate the problem of engineering synchronization in non-Markovian quantum systems. First, a time-convoluted linear quantum stochastic differential equation is derived which describes the Heisenberg evolution of a localized quantum system driven by multiple colored noise inputs. Then, we define quantum expectation synchronization in an augmented system consisting of two subsystems. We prove that, for two homogenous subsystems, synchronization can always be synthesized without designing direct Hamiltonian coupling given that the degree of non-Markovianity is below a certain threshold. System parameters are explicitly designed to achieve quantum  synchronization. Also, a numerical example is presented to illustrate our results.      
\end{abstract}

\end{frontmatter}

\section{Introduction}
In the late 17th century, Huygens noticed the gradual alignment of motion between two pendulum clocks hung from a common beam. This was the prologue to the study of synchronization, which, in general terms, refers to the asymptotic unification of dynamics in different subsystems due to mutual coupling or external force. Actually, this is a phenomenon ubiquitously observed in biological, chemical, social and many other complex classical systems \cite{JAFARIZADEH2020108711,9022900}.

Not surprisingly, the concept of synchronization has been extended to the quantum domain, where dynamics of physical systems are governed by quantum mechanics laws fundamentally different from those in the classical regime. Studying how synchronization manifests itself in quantum systems is of great theoretical significance, since, on one hand, it can deepen our understanding of collective quantum dynamics and many-body quantum systems \cite{PhysRevA.90.033603}, and on the other hand, it has promising applications in quantum information science, e.g., synchronizing remote nodes in quantum communication networks \cite{PhysRevLett.125.013601}.

Existing literature regarding quantum synchronization has covered a wide range of physical systems, both finite dimensional  \cite{PhysRevLett.125.013601,PhysRevLett.121.063601,PhysRevLett.123.023604,PhysRevE.101.020201,TICOZZI201638,JAFARIZADEH2016237,ALBERTINI201655,7907189,7109119,6849451} and Continuous-Variable (CV) \cite{PhysRevLett.120.163601,PhysRevA.94.052118,PhysRevLett.118.243602,PhysRevLett.117.073601}, with the majority of them considering closed systems or Markovian open systems. However, to the best of our knowledge, few works have discussed synchronization problems of non-Markovian quantum systems. In many realistic scenarios, the gap between system and environment time scales is simply not large enough to meet the Markovian assumptions \cite{RevModPhys.89.015001}. Consequently, non-Markovian effects in quantum systems naturally arise, leading to new dynamical features that may complicate the design of quantum information processing structures and bring potential resource to harness \cite{PhysRevLett.109.233601}. Non-Markovianity of quantum systems has significant implications to unique quantum resources such as coherence and non-classical correlation and investigation of non-Markovian quantum systems has attracted wide attention in recent years \cite{RevModPhys.89.015001,LI20181,2020Detecting,PhysRevA.101.042327,8820133,PhysRevA.77.032117}. In this context, we are motivated to investigate the synchronzation problem of non-Markovian quantum systems. 

Admittedly, the scope of non-Markovian systems itself is too vast to be covered in a single work. Here, we are particularly interested in quantum systems describable by the input-output formalism \cite{PhysRevA.31.3761}, a notion which provides convenience for the application of control theory to quantum physics. The input-output formalism was originally considered under Markovian assumption, which leads to Quantum Stochastic Differential Equations (QSDEs) depicting the Heisenberg evolution of localized systems driven by quantum white noise. In the case where system variables are observables satisfying Canonical Commutation Relations (CCR) and the Hamitonian is quadratic and the coupling operators are linear in those variables, linear QSDEs can be obtained to describe the system dynamics. These systems are termed as ``linear quantum systems" and can be well described by quantum stochastic calculus \cite{1984Quantum}. In the past two decades, this formalism has nurtured great success in the study of linear quantum systems and coherent feedback control \cite{4625217,NURDIN20091837,6777569,GRIVOPOULOS2018103,8862949,LEVITT2018255}.  

The input-output formalism has been extended to non-Markovian systems \cite{PhysRevA.85.034101,PhysRevA.87.032117}.  In \cite{PhysRevA.87.032117}, non-Markovian QSDEs were derived where colored noise replaces white noise in the Markovian scenario. Also, a time-convolution term is present for containing all historical system-environment interactions. Building on similar derivation as in \cite{PhysRevA.87.032117}, if we again consider a localized system with quadratic Hamiltonian, linear couplings and variables satisfying CCR, we end up with a \textbf{linear, time-convoluted QSDE driven by colored noise input}: a non-Markovian linear quantum system. Such a QSDE directly allows for the study of synchronization in the sense of quantum expectation, which is defined by the asymptotically vanishing gap between the expectations of individual subsystems' states regardless of the initial quantum state of the augmented system.  

More specifically, we would like to answer the following questions: 1) Given a non-Markovian linear quantum system with the inherent parameters of its two subsystems fixed, under what conditions would it be possible for the augmented system to achieve synchonization? 2) Provided that those conditions are met, how to construct the remaining parameters to achieve synchronization? The purpose of studying the problems stated above is two-fold. Firstly, on a theoretical note, we explore the interplay between synchronization, system stability and quantum dynamics with memory. Secondly, on practical grounds, our findings may lead to new applications in engineering quantum devices where non-Markovian effects cannot be neglected.

The rest of this paper is organized as follows. In Section~ 2, we present how non-Markovian linear QSDEs are derived from a microscopic physical model. In Section~ 3, the problem of synchronization in quantum expectation sense is defined, which is followed by the derivation of synchronization error evolution equation and the analysis on the conditions for achieving synchronization. Section 4 includes a numerical example to illustrate our results. In Section 5, we conclude our paper and briefly discuss possible future directions.

\textbf{Notation.} In this work, Roman type character i represents the imaginary unit while italic type $i$ is used for indexing. Let $A$ , $B$ be arbitrary matrices and $\hat{A}$ , $\hat{B}$ be Hilbert space operators. We denote by $A^\dagger$ the conjugate transpose of $A$ and $A^T$ the transpose of $A$. $\text{Re}(A)$ and $\text{Im}(A)$ denote the real and imaginary parts of $A$, respectively. $\text{Det}(A)$ stands for the determinant of $A$. $\text{Ker}(A)$ is the kernel space of $A$. We specify that $\Vert A \Vert$ denotes the 2-norm of $A$. $A \otimes B$ is the Kronecker product of $A$ and $B$. $\hat{A}^\dagger$ is the adjoint operator of $\hat{A}$. $\text{tr}(\hat{A})$ is the trace of $\hat{A}$. We denote the quantum expectation of $\hat{A}$ by $\langle \hat{A}\rangle$. $[\hat{A},\hat{B}]=\hat{A}\hat{B}-\hat{B}\hat{A}$ denotes the commutator of $\hat{A}$, $\hat{B}$ and $\hat{A} \otimes \hat{B}$ represents their tensor product. Moreover, for positive integers $m$, $n$, $I_n$ is the $n \times n$ identity matrix and $0_{m \times n}$ is the $m \times n$ zero matrix. $\mathbb{R}^{n}$ and $\mathbb{C}^{n}$ denote real space and complex space of dimension $n$ respectively. $|\psi\rangle$ and $\langle\phi |$ are Dirac representation of quantum states.




\section{System Dynamics}
In this section, the non-Markovian quantum system on which our synchronization study is based is derived from a microscopic model. The mathematical model is formulated via a generalization of the non-Markovian QSDE in \cite{PhysRevA.87.032117} to multiple colored noise input scenario, with variables satisfying CCR, quadratic Hamiltonian and linear couplings.

\subsection{QSDE from a Microscopic Model}
The physical model we consider is a localized system with Hamiltonian $H_S$ interacting with $M$ bosonic fields, which are modelled as infinite collections of harmonic oscillators comprising of a continuum of frequencies. The annihilation operators of the bosonic fields, $b_1(\omega),...,b_M(\omega)$, satisfy the following commutation relation:
\begin{equation}
[b_i(\omega),b_{j}^{\dagger}(\tilde{\omega})]=\delta_{ij}\delta(\omega-\tilde{\omega}), \quad 1\leq i,j \leq M.
\end{equation}
Here, $\delta_{ij}=1$ if $i=j$, and $\delta_{ij}=0$ if otherwise. Also, $\delta(\cdot)$ represents Dirac delta function.

The $M$ bosonic fields serve as the bath of our system, with bath Hamiltonian expressed as:
\begin{equation}
H_B=\sum_{j=1}^{M}\int_{-\infty}^{+\infty}\omega b_{j}^{\dagger}(\omega)b_{j}(\omega) d\omega .
\end{equation}

The interaction interface between the system and fields are $M$ coupling operators acting on the system's Hilbert space: $\bm{L}=(L_1,...,L_M)^T$. The interaction Hamiltonian is written as:
\begin{equation}
H_{\text{int}}=\text{i}\sum_{j=1}^{M}\int_{-\infty}^{+\infty}\Big{(}\kappa_{j}(\omega) b_{j}^{\dagger}(\omega)L_j-\kappa_{j}^{*}(\omega) b_{j}(\omega)L_j^{\dagger}\Big{)}d\omega,
\end{equation}
where $\kappa_{j}(\omega),1 \leq  j \leq M$, represent frequency coupling strength and are assumed to be constant in the Markovian case \cite{PhysRevA.31.3761}.  In this article, we assume that for $1\leq j \leq M$, $\kappa_{j}(\omega)$ is real.

Next, the system and bath are treated together as a closed system, which leads to the evolution of an arbitrary system operator $A$ in the Heisenberg picture (set $\hbar=1$):
\begin{equation}
\dot{A}=-\text{i}[A, H_S+H_B+H_{\text{int}}].
\end{equation}

Following the same procedure as presented in \cite{PhysRevA.31.3761,PhysRevA.87.032117} while keeping in mind that we are now concerned with $M$ bosonic fields instead of only one, we arrive at the following non-Markovian QSDE describing the Heisenberg evolution of $A$:
\begin{multline}
\dot{A}=-\text{i}[A, H_S]+\bm{\tilde{b}_\text{in}}^{\dagger}[A,\bm{L}]+[\bm{L}^{\dagger},A]\bm{\tilde{b}_\text{in}}\\
+\int_0^t \Big{(}\bm{L}^{\dagger}(\tau) \Gamma^{\dagger}(t-\tau)[A,\bm{L}]+[\bm{L}^{\dagger},A]\Gamma(t-\tau)\bm{L}(\tau)\Big{)} d\tau.
\end{multline}
In (5), $\bm{\tilde{b}_\text{in}}=(\tilde{b}_{\text{in},1},...,\tilde{b}_{\text{in},M})^T$ is a vector of colored noise input, with each of its component written as ($1 \leq j \leq M$):
\begin{align}
&\tilde{b}_{\text{in},j}(t)=\int_{-\infty}^{+\infty}\kappa_j(\omega)e^{-\text{i}\omega t}b_j(\omega)d\omega
\end{align}
Moreover, $\Gamma(t)$ is a diagonal $M \times M$ matrix function, i.e., $\Gamma(t)=\text{diag}(\gamma_1(t),...,\gamma_M(t))$, where
\begin{equation}
\gamma_j(t)=\int_{-\infty}^{+\infty}e^{-\text{i}\omega t}\kappa_j^2(\omega)d\omega.
\end{equation}
Clearly, (5) indicates that, in the absence of scattering, it takes three parameters to pin down the Heisenberg evolution of a localized system driven by colored noise: 1) system Hamiltonian $H_S$ for internal energy; 2) a vector of coupling operaters $\bm{L}=(L_1,...,L_M)^T$ describing the interaction interface; 3) a memory kernel matrix function $\Gamma(t)$ weighing all history environmental interactions. 

It is worthwhile comparing (5) to its Markovian counterpart:
\begin{align}
\dot{A}=&-\text{i}[A, H_S]+\bm{b_\text{in}}^{\dagger}[A,\bm{L}]+[\bm{L}^{\dagger},A]\bm{b_\text{in}}\\ \nonumber
&+\bm{L}^{\dagger}[A,\bm{L}]+[\bm{L}^{\dagger},A]\bm{L},
\end{align}
where $\bm{b_\text{in}}=(b_{\text{in},1},...,b_{\text{in},M})^T$. It can be checked that there are two major differences. Firstly, Markovian systems (8) are driven by white noise satisfying commutation relation
\begin{equation}
[b_{\text{in},i}(t),b_{\text{in},j}^{\dagger}(\tilde{t})]=\delta_{ij}\delta(t-\tilde{t}), \quad 1\leq i,j \leq M,
\end{equation}
while non-Markovian systems (5) are driven by colored noise which satisfies:
\begin{equation}
[\tilde{b}_{\text{in},i}(t),\tilde{b}_{\text{in},j}^{\dagger}(\tilde{t})]=\delta_{ij}\gamma_j(t-\tilde{t}), \quad 1\leq i,j \leq M.
\end{equation}
Secondly, in the Markovian case, system evolution at time $t$ depends only on current conditions, whereas non-Markovian systems take into account all history conditions.

\subsection{Non-Markovian Linear QSDE}
In fact, (5) describes a quite general class of open quantum systems. It is well-known, in the Markovian case, that if system variables are canonical variables satisfying CCR, $H_S$ is quadratic and $\bm{L}$ is linear in system variables, a set of linear QSDE can be derived. Now, applying the same restrictions to (5), it can be checked that we will also end up with linear QSDE. However, what is new is time convolution terms and colored noise inputs signifying non-Markovianity. 

Consider a system in equivalence with $n$ quantum harmonic oscillators, whose variables are $2n$ canonical operators: $x=(x_{1},...,x_{2n})^T$. These operators satisfy the following CCR:
\begin{equation}
[x_j,x_k]=(\text{i}J_n)_{jk},\quad 1\leq j,k \leq 2n,
\end{equation}
where $J_n=I_n \otimes J$, and $J=\begin{pmatrix} 0 & 1 \\ -1 & 0 \end{pmatrix}$. Meanwhile, let system Hamiltonian be expressed as $H_S=x^T \Omega x$, where $\Omega \in \mathbb{R}^{2n \times 2n}$ is a symmetric matrix. Moreover, the coupling operators take linear form: $\bm{L}=Cx$, where $C \in \mathbb{C}^{M \times 2n}$.

Next, by repeatedly applying (11), the subsitution of above defined $x$, $H_S$ and $\bm{L}$ into (5) yields the following non-Markovian linear QSDE:
\begin{equation}
\dot{x}=A_H x+\int_{0}^{t}A_K(t-\tau)x(\tau)d\tau+B\tilde{b},
\end{equation}
where $A_H=2J_n \Omega$, $B=\text{i}J_n(-C^{\dagger}\quad C^T)$, $A_K(t)=2J_n \text{Im}(C^{\dagger}\Gamma(t)C)$ and $\tilde{b}=(\bm{\tilde{b}_\text{in}},\bm{\tilde{b}_\text{in}}^{\dagger})^T$.

The equation above indicates that the tuple $(\Omega, C, \Gamma(t))$ completely specifies a non-Markovian linear quantum system. Moreover, since (12) is derived directly from microscopic model based on first principle, it not only represents a mathematical entity, but also describes certain aspects of physical reality. Our synchronization study is based on system (12).

\section{Quantum Synchronization}
In this section, the problem of quantum synchronization is formulated. We derive autonomous evolution equation of synchonization error, based on which we are able to present sufficient conditions for synchronization, which is then achieved by explicitly designing free system parameters.

\subsection{Problem Statement}
Consider two non-Markovian linear quantum systems specified by $(\Omega_j, V_j, \Gamma_j(t))$, whose variables are denoted by $\xi_j$, $j=1,2$. Here, $\Omega_j \in \mathbb{R}^{2n \times 2n}$ is symmetric and $V_j \in \mathbb{C}^{M \times 2n}$. We assume that all diagonal entries of $\Gamma_j(t)$ are real and positive for $t \geq 0$, which corresponds to the physical interpretation of memory kernel. Another assumption is that $\Vert \Gamma_j(t) \Vert \in L^1 [0,\infty)$. 

The $M$ in the size of matrices reflects the number of coupling operators in the system, or equivalently, the number of bosonic fields the system interacts with. In the event that $M<n$, the system can also be viewed as being bathed in $n$ bosonic fields but only being in contact with $M$ of them. Therefore, after introducing $n-M$ dummy input fields, augmenting $V_j$ to $\begin{pmatrix} V_j \\ 0_{(n-M) \times 2n} \end{pmatrix}$ and $\Gamma_j(t)$ to $\begin{pmatrix} \Gamma_j(t) & 0 \\ 0 & \tilde{\Gamma}_j(t)\end{pmatrix}$, the dynamics of two systems are unchanged. Therefore, in this work, we further assume that $M \geq n$. 

Firstly, the two systems are treated separately with their dynamics uncorrelated. To proceed, we put the two systems in contact to form an augmented system, which consequently demotes both of them to subsystems. Synchronization will then be defined and studied with respect to the augmented system.

We build the augmented system in the way that it is still describable via non-Markovian linear QSDE. Therefore, the augmented system is still pinned down by a 3-tuple, where each subsystem retains the parameters of $(\Omega_j, V_j, \Gamma_j(t))$, $j=1,2$, and the augmented system contains new Hamiltonian interactions and coupling operators to be engineered. Let us denote the variables of the augmented system by $\xi=(\xi_1,\xi_2)^T$, where $\xi_j$, $j=1,2$ are the variables of the constituent subsystems. The tuple $(R,V,\Gamma(t))$ that pins down the augmented system takes the following form:
\begin{equation}
R=\begin{pmatrix} \Omega_1 & \Omega_{12} \\ \Omega_{12}^T & \Omega_2 \end{pmatrix},\quad V=\begin{pmatrix} V_1 & V_{12} \\ V_{21} & V_2 \end{pmatrix},
\end{equation}
and
\begin{equation}
\Gamma(t)=\begin{pmatrix} \Gamma_1(t) & 0 \\ 0 & \Gamma_2(t)\end{pmatrix}.
\end{equation}
The QSDE for the augmented system is expressed as
\begin{equation}
\dot{\xi}=\mathcal{A}_H \xi + \int_{0}^{t}\mathcal{A}_K(t-\tau)\xi(\tau)d\tau+\mathcal{B}\tilde{b}_{\text{aug}},
\end{equation}
where $\mathcal{A}_H=2J_{2n}R$, $\mathcal{B}=\text{i}J_{2n}(-V^{\dagger}\quad V^T)$, $\mathcal{A}_K(t)=2J_{2n} \text{Im}(V^{\dagger}\Gamma(t)V)$ and $\tilde{b}_{\text{aug}}=(\tilde{b}_{\text{in},1},\tilde{b}_{\text{in},2},\tilde{b}_{\text{in},1}^{\dagger},\tilde{b}_{\text{in},2}^{\dagger})^T$.

Provided this formal expression, we give the following definition pertaining to the synchronization problem under study.

{\bfseries Definition 1.} \textit{System (15) is said to achieve sychronization in quantum expectation sense, if for any intial state with form} $\rho_{\text{ini}}=\rho_{\text{sys}} \otimes |0\rangle  \langle 0|$ \textit{(product of an arbitrary system state with vacuum field state), the following conditions hold:}

\textit{(C1)} $\langle \xi_1(0) \rangle= \langle \xi_2(0) \rangle$ \textit{implies} $\langle \xi_1(t) \rangle= \langle \xi_2(t) \rangle$, $\forall t \geq 0$;

\textit{(C2)} $\lim\limits_{t \to \infty} \langle \xi_1(t)\rangle- \langle \xi_2(t) \rangle=0$.

It is clear from (13) and (14) that $(\Omega_j, V_j, \Gamma_j(t))$, $j=1,2$ are original parameters for the two subsystems while $\Omega_{12}$, $V_{12}$ and $V_{21}$ are new parameters to be engineered for the augmented system. 
The goal of our study is \textbf{1) to find a set of conditions on system parameters under which sychronization in quantum expectation sense can be engineered, and 2) to properly choose $\Omega_{12}$, $V_{12}$ and $V_{21}$ so that sychronization is reached.}

Note that, to the best of our knowledge, formal results on physical realizability \cite{4625217} of non-Markovian linear quantum systems have not been developed. Therefore, after working with $\mathcal{A}_H$, $\mathcal{A}_K(t)$ and $\mathcal{B}$, our restrictions are ultimately casted on parameters defining system Hamilonian and coupling operators so that physical realizability is automatically satisfied. Moreover, these parameters are directly related to experimentally engineering the system dynamics.

\subsection{Evolution of Synchronization Error}
Since it is required in Definition 1 that the initial state be a separable state where the fields are in vacuum, taking quantum expectation with respect to the initial state on both sides of (15) yields:
\begin{equation}
\langle \dot{\xi} \rangle=\mathcal{A}_H \langle \xi \rangle + \int_{0}^{t}\mathcal{A}_K(t-\tau) \langle \xi(\tau) \rangle d\tau.
\end{equation}
Let us introduce a matrix $\mathscr{G}=(-I_{2n}\quad I_{2n})$ and denote by $\Pi$ the orthogonal projection operator from $\mathbb{R}^{4n}$ to $\text{Ker}(\mathscr{G})$. The conditions in Definition 1 can be rephrased as $\mathscr{G}\langle \xi(0) \rangle=0$ implies $\mathscr{G}\langle \xi(t) \rangle=0$, $\forall t \geq 0$ and $\lim\limits_{t \to \infty} \mathscr{G}\langle \xi(t) \rangle=0$. 

Let $\Pi^\perp$ be the orthogonal complement of  $\Pi$. It can be checked that $\mathscr{G}\langle \xi \rangle=\mathscr{G} \Pi^\perp \langle \xi \rangle$. Also, $\mathscr{G}\langle \xi \rangle=0$ if and only if $\Pi^\perp \langle \xi \rangle=0$. In light of this, the two conditions in Definition 1 can be further equivalently translated into:

(C1') $\Pi^\perp  \langle \xi(0) \rangle=0 \Rightarrow \Pi^\perp  \langle \xi(t) \rangle=0$, $\forall t \geq 0$;

(C2') $\lim\limits_{t \to \infty} \Pi^\perp \langle \xi(t) \rangle=0$.

The conditions above suggests that, to achieve synchronization in quantum expectation sense, parameters of the augmented system must be designed such that $\text{Ker}(\mathscr{G})$ is an invariant and attractive subspace of $\mathbb{R}^{4n}$ in terms of dynamics generated by (16). Moreover, $\Pi^\perp \langle \xi \rangle$, which is the component outside $\text{Ker}(\mathscr{G})$, should be kept at zero or driven infinitely close to zero.

At this point, it is helpful to derive separate equations depicting the evolution of $\Pi^\perp \langle \xi \rangle$ and $\Pi \langle \xi \rangle$ from (16). Taking advantage of the idempotence of $\Pi^\perp$ and $\Pi$, we have
\begin{align}
\dot{\Pi}\langle \xi \rangle=&(\Pi \mathcal{A}_H \Pi)(\Pi \langle \xi \rangle) \\ \nonumber
                                          &\!+(\Pi \mathcal{A}_H \Pi^\perp)(\Pi ^\perp \langle \xi \rangle) \\ \nonumber
                                          &\!+\int_{0}^{t}(\Pi \mathcal{A}_K(t-\tau) \Pi)(\Pi \langle \xi(\tau) \rangle) d\tau \\ \nonumber
                                          &\!+\int_{0}^{t}(\Pi \mathcal{A}_K(t-\tau) \Pi^\perp)(\Pi^\perp \langle \xi(\tau) \rangle) d\tau ,
\end{align}
and
\begin{align}
\dot{\Pi}^\perp \langle \xi \rangle=&(\Pi^\perp \mathcal{A}_H \Pi)(\Pi \langle \xi \rangle) \\ \nonumber
                                                  &\!+(\Pi^\perp \mathcal{A}_H \Pi^\perp)(\Pi ^\perp \langle \xi \rangle) \\ \nonumber
                                                  &\!+\int_{0}^{t}(\Pi^\perp \mathcal{A}_K(t-\tau) \Pi)(\Pi \langle \xi(\tau) \rangle) d\tau \\ \nonumber
                                                  &\!+\int_{0}^{t}(\Pi^\perp \mathcal{A}_K(t-\tau) \Pi^\perp)(\Pi^\perp \langle \xi(\tau) \rangle) d\tau .
\end{align}

Given (17) and (18), we present the following results.

\begin{thm}
A sufficient condition of achieving (C1) is $\Pi^\perp \mathcal{A}_H \Pi=0$ and $\Pi^\perp \mathcal{A}_K(t) \Pi=0$, $\forall t \geq 0$. Meanwhile, $\Pi^\perp \mathcal{A}_H \Pi=0$ is necessary for (C1) to hold.
\end{thm}  

\begin{pf}
Since (C1) and (C1') are equivalent, it suffices to show that $\Pi^\perp \mathcal{A}_H \Pi=0$ and $\Pi^\perp \mathcal{A}_K(t) \Pi=0$, $\forall t \geq 0$ lead to (C1'), and that (C1') entails $\Pi^\perp \mathcal{A}_H \Pi=0$.

If $\Pi^\perp \mathcal{A}_H \Pi=0$ and $\Pi^\perp \mathcal{A}_K(t) \Pi=0$, $\forall t \geq 0$, (18) is reduced to:
\begin{align}
\dot{\Pi}^\perp \langle \xi \rangle&=(\Pi^\perp \mathcal{A}_H \Pi^\perp)(\Pi ^\perp \langle \xi \rangle) \\ \nonumber
                                                   &+\int_{0}^{t}(\Pi^\perp \mathcal{A}_K(t-\tau) \Pi^\perp)(\Pi^\perp \langle \xi(\tau) \rangle) d\tau .                                     
\end{align}
Apparently, $\Pi^\perp \langle \xi \rangle \equiv 0$ is a solution, and the only solution with initial state 0, to (19). Therefore, if $\Pi^\perp  \langle \xi(0) \rangle=0$, we must have $\Pi^\perp  \langle \xi(t) \rangle=0$, $\forall t \geq 0$. (C1') thus follows.

On the other hand, suppose that (C1') holds, then $(\Pi  \langle \xi \rangle , 0)$ is a solution to the closed-form equation combining (17) and (18). The following equations thus hold:
\begin{align}
\dot{\Pi}\langle \xi \rangle=&(\Pi \mathcal{A}_H \Pi)(\Pi \langle \xi \rangle) \\ \nonumber
                                          &\!+\int_{0}^{t}(\Pi \mathcal{A}_K(t-\tau) \Pi)(\Pi \langle \xi(\tau) \rangle) d\tau ; 
\end{align}
\begin{align}
0=&(\Pi^\perp \mathcal{A}_H \Pi)(\Pi \langle \xi \rangle) \\ \nonumber
  &\!+\int_{0}^{t}(\Pi^\perp \mathcal{A}_K(t-\tau) \Pi)(\Pi \langle \xi(\tau) \rangle) d\tau .
\end{align}

In (21), setting $t=0$ yields $\Pi^\perp \mathcal{A}_H \Pi \langle \xi(0) \rangle=0$ for arbitrary $\langle \xi(0) \rangle$. To proceed, we shall constructively prove that the choice of system initial state is abundant enough so that all possible values of $\langle \xi(0) \rangle$ span the entire $\mathbb{R}^{4n}$.

Our construction builds on the following calculation. Consider a single node coherent state $| \alpha \rangle$, $\alpha \in \mathbb{C}$. Denoting position and momentum operators by $q$ and $p$, respectively, we have
\begin{equation}
\langle \alpha | q | \alpha \rangle=\sqrt{2}\text{Re}(\alpha),\quad \langle \alpha | p | \alpha \rangle=\sqrt{2}\text{Im}(\alpha).
\end{equation}

Building on (22), we construct the following family of multi-mode initial states:
\begin{equation}
\rho_{q,k}=|0\rangle \langle 0|^{\otimes^{k-1}}\otimes |\theta \rangle \langle \theta |\otimes |0\rangle \langle 0|^{\otimes^{2n-k}}, 1\leq k \leq 2n,
\end{equation}

\begin{equation}
\rho_{p,k}=|0\rangle \langle 0|^{\otimes^{k-1}}\otimes |\theta \text{i} \rangle \langle \theta \text{i} |\otimes |0 \rangle \langle 0 |^{\otimes^{2n-k}}, 1\leq k \leq 2n,
\end{equation}
where $\theta \in \mathbb{R}$ and $\theta \neq 0$.

For $1 \leq j,k \leq 2n$, the initial states (23) and (24) lead to the following quantum expectations:
\begin{align}
&\text{tr}(\xi_{2j-1}\rho_{q,k}) =\sqrt{2}\delta_{jk}\theta, \\ \nonumber
&\text{tr}(\xi_{2j}\rho_{q,k}) =0, \\ \nonumber
&\text{tr}(\xi_{2j-1}\rho_{p,k}) =0, \\ \nonumber
&\text{tr}(\xi_{2j}\rho_{p,k}) =\sqrt{2}\delta_{jk}\theta.
\end{align}
It is now clear from (25) that 
\begin{align*}
\mathbb{R}^{4n}=\text{span}\{\text{tr}(\rho_{r,k}\xi)|r=p,q;1 \leq k \leq 2n \}.
\end{align*}
This indicates that the arbitrariness of $\xi(0)$ extends to $\mathbb{R}^{4n}$, which further leads to $\Pi^\perp \mathcal{A}_H \Pi=0$. 
\end{pf}
Next, the requirements on QSDE matrices in Theorem 1 are transplanted to that on parameters in system Hamiltonian and coupling operators with which we are ultimately concerned. To begin with, the orthogonal projections admit the following matrix representations:
\begin{equation}
\Pi=\frac{1}{2}\begin{pmatrix} I_{2n} & I_{2n} \\ I_{2n} & I_{2n} \end{pmatrix},\quad \Pi^{\perp}=\frac{1}{2}\begin{pmatrix} I_{2n} & -I_{2n} \\ -I_{2n} & I_{2n} \end{pmatrix}.
\end{equation}
 In fact, when a given matrix is acted upon by these projections, certain block structures will arise. Let $D=\begin{pmatrix} D_{11} & D_{12} \\ D_{21} & D_{22} \end{pmatrix}$ be an arbitrary $4n \times 4n$ matrix, with all subblocks in size $2n \times 2n$. Then, we have
\begin{equation}
\Pi^\perp D \Pi=\frac{1}{4}\begin{pmatrix} D_1 & D_1 \\ -D_1 & -D_1 \end{pmatrix},
\end{equation}
where
\begin{equation}
D_1=D_{11}-D_{21}+D_{12}-D_{22}.
\end{equation}
With the aid of (27) and (28), the restrictions in Theorem 1 can be conveniently casted on system Hamiltonian and coupling operators, which is encapsulated in the following corollary.

{\bfseries Corollary 1.} \textit{Consider the following restrictions on system Hamiltonian and coupling operators:}
\begin{equation}
\Omega_1+\Omega_{12}=\Omega_{12}^T+\Omega_2,
\end{equation}

\begin{align}
&\text{Im}(V_{1}^\dagger \Gamma_1(t)V_{1}+V_{21}^\dagger \Gamma_2(t)V_{21}) \\ \nonumber
&+\text{Im}(V_{1}^\dagger \Gamma_1(t)V_{12}+V_{21}^\dagger \Gamma_2(t)V_{2}) \\ \nonumber
&=\text{Im}(V_{12}^\dagger \Gamma_1(t)V_{1}+V_{2}^\dagger \Gamma_2(t)V_{21}) \\ \nonumber
&+\text{Im}(V_{12}^\dagger \Gamma_1(t)V_{12}+V_{2}^\dagger \Gamma_2(t)V_{2}),
\end{align}


\begin{equation}
\Omega_1=\Omega_2,\quad \Omega_{12}=\Omega_{12}^T.
\end{equation}
\textit{In terms of (C1), (29) and (30) are sufficient while (29) and (31) are necessary.}

\begin{pf}
The QSDE matrices admit the following subblock structures:
\begin{equation}
\mathcal{A}_H=2\begin{pmatrix} J_n\Omega_{1} & J_n\Omega_{12} \\  J_n\Omega_{12}^T &  J_n\Omega_{2} \end{pmatrix}, \mathcal{A}_K(t)=\begin{pmatrix} \mathcal{A}_K^{11}(t) & \mathcal{A}_K^{12}(t) \\  \mathcal{A}_K^{21}(t) &  \mathcal{A}_K^{22}(t) \end{pmatrix}, 
\end{equation}
where
\begin{align}
&\mathcal{A}_K^{11}(t)=2J_n\text{Im}(V_{1}^\dagger \Gamma_1(t)V_{1}+V_{21}^\dagger \Gamma_2(t)V_{21}),  \\ \nonumber
&\mathcal{A}_K^{12}(t)=2J_n\text{Im}(V_{1}^\dagger \Gamma_1(t)V_{12}+V_{21}^\dagger \Gamma_2(t)V_{2}),  \\ \nonumber
&\mathcal{A}_K^{21}(t)=2J_n\text{Im}(V_{12}^\dagger \Gamma_1(t)V_{1}+V_{2}^\dagger \Gamma_2(t)V_{21}),  \\ \nonumber
&\mathcal{A}_K^{22}(t)=2J_n\text{Im}(V_{12}^\dagger \Gamma_1(t)V_{12}+V_{2}^\dagger \Gamma_2(t)V_{2}).
\end{align}
Given these explicit expressions, we first prove the sufficiency of (29) and (30).  Suppose that (29) and (30) hold, by substituting (32) and (33) into (27) and (28), it can be directly checked that $\Pi^\perp \mathcal{A}_H \Pi=0$ and $\Pi^\perp \mathcal{A}_K(t) \Pi=0$, $\forall t \geq 0$. This consequently leads to (C1) according to Theorem 1. 

For necessity of (29) and (31), Theorem 1 along with (27), (28) and (32) yields
\begin{equation}
J_n(\Omega_1+\Omega_{12})=J_n(\Omega_{12}^T+\Omega_2).
\end{equation}
Due to the invertibility of $J_n$, (29) thus follows. Next, (29) says that
\begin{equation}
\Omega_1-\Omega_2=\Omega_{12}^T-\Omega_{12}.
\end{equation}
Since $\Omega_1$ and $\Omega_2$ are symmetric, performing matrix transpose on both sides of (35) yields
\begin{equation}
\Omega_1-\Omega_2=\Omega_{12}-\Omega_{12}^T.
\end{equation}
Eqs. (35) and (36) necessarily entail that $\Omega_{12}=\Omega_{12}^T$ and $\Omega_1=\Omega_2$. Eq. (31) thus follows.
\end{pf}

{\bfseries Remark 1.} \textit{In Corollary 1, (29) and (30) are only equivalent with their corresponding statements in Theorem 1. However, (31) presents new restrictions which are not covered by Theorem 1. The restrictions are resulted from the symmetry of system Hamiltonian. Specifically speaking, quantum synchronization in Definition 1 \textbf{cannot} occur if two subsystems carry different inherent Hamiltonians. Moreover, engineered blocks in the augmented Hamiltonian must be symmetric as well.}

It is noticable that the sufficient conditions proposed in Theorem 1 and Corollary 1 not only guarantee (C1), but also enable the independent evolution of $\Pi^\perp \langle \xi \rangle$, as described by (19). Clearly, we have
\begin{equation}
\Pi^\perp \langle \xi \rangle=\begin{pmatrix}  \langle \xi_1 \rangle-\langle \xi_2 \rangle \\ \langle \xi_2 \rangle-\langle \xi_1 \rangle \end{pmatrix},
\end{equation}
where $e \triangleq \langle \xi_1 \rangle-\langle \xi_2 \rangle$ is exactly the synchronization error we are concerned with. This motivates us to derive an independent equation for $e$ as well. 

Provided that $\Pi^\perp D \Pi=0$, it can be checked that
\begin{equation}
\Pi^\perp D \Pi^\perp=\frac{1}{2}\begin{pmatrix} D_2 & -D_2 \\ -D_2 & D_2 \end{pmatrix},
\end{equation}
where
\begin{equation}
D_2=D_{11}-D_{21}.
\end{equation}
Eqs. (38) and (39) significantly compress the derivation of the evolution of $e$, which will be shown in the following corollary.

{\bfseries Corollary 2.} \textit{Suppose that (29) and (30) hold, synchronization error evolves according to the following equation:}
\begin{equation}
\dot{e}=Ee+\int_0^t F(t-\tau)e(\tau)d\tau,
\end{equation}
where
\begin{equation}
E=2J_n (\Omega_1-\Omega_{12}^T),
\end{equation}
and
\begin{align}
F(t)=&2J_n\text{Im}(V_{1}^\dagger \Gamma_1(t)V_{1}+V_{21}^\dagger \Gamma_2(t)V_{21})\\ \nonumber
        &\!-\!2J_n\text{Im}(V_{12}^\dagger \Gamma_1(t)V_{1}+V_{2}^\dagger \Gamma_2(t)V_{21}). 
\end{align}

\begin{pf}
The fact that (29) and (30) hold indicates that (19) holds. The results thus follow by substituting (32) and (37) into (19), (38) and (39).
\end{pf}

\subsection{Asymptotic Stability of Synchronization Error}
In the previous subsection, we have specified some sufficient conditions that lead to (C1) and independent evolution of synchronization error. Building on these conditions, we investigate in this subsection whether (C2) can be satisfied, or in other words, whether system parameters can be designed so as to make the zero solution of (40) asymptotically stable. 

To proceed, a lemma is proved which is helpful to the derivation of our main result.

{\bfseries Lemma 1.} \textit{Consider the following dynamical system}
\begin{equation}
\dot{y}=\mathcal{E}y+\int_0^t \mathcal{F}(t-\tau)y(\tau)d\tau,
\end{equation}
\textit{where} $\mathcal{E}\in \mathbb{R}^{N\times N}$, $\mathcal{F}(t): [0,\infty) \to \mathbb{R}^{N\times N}$ \textit{is continuous and} $\Vert \mathcal{F}(t)\Vert \in L^1[0,\infty)$.  \textit{Suppose that} $\mathcal{E}+\int_0^\infty \mathcal{F}(t)dt$ \textit{is Hurwitz, and then the zero solution of (43) is asymptotically stable if the following conditions are satisfied:}

\textit{(C3)} $\int_0^\infty t\Vert \mathcal{F}(t)\Vert dt <\infty$;

\textit{(C4)} \textit{Let} $\mathcal{F}_+ \triangleq \int_0^\infty \Vert \mathcal{F}(t)\Vert dt$ \textit{and} $p_{\mathcal{F}}(t) \triangleq \Vert \mathcal{F}(t)\Vert/\mathcal{F}_+$. \textit{Then}
\begin{equation}
\int_0^\infty t p_{\mathcal{F}}(t)dt <\frac{2|\text{Re}(\lambda_1)|^N}{N(2\Vert \mathcal{E}\Vert+2\mathcal{F}_+)^N \mathcal{F}_+ },
\end{equation}
\textit{where} $\lambda_1$ \textit{denotes the eigenvalue of} $\mathcal{E}+\int_0^\infty \mathcal{F}(t)dt$ \textit{with the largest real part.}
\begin{pf}
Let $\hat{\mathcal{F}}(s)$ and $\hat{y}(s)$  denote the Laplace transform of $\mathcal{F}(t)$ and $y(t)$. By performing Laplace transform on both sides of (43) and collecting similar terms, we obtain
\begin{equation}
(sI_N-\mathcal{E}-\hat{\mathcal{F}}(s))\hat{y}(s)=y(0).
\end{equation}
We shall prove that $\text{Det}(sI_N-\mathcal{E}-\hat{\mathcal{F}}(s))\neq 0$ for all $s$ on the right half complex plane and the imaginary axis, i.e., $\text{Re}(s)\geq 0$.

The case where $|s|>\Vert \mathcal{E}\Vert+\mathcal{F}_+$ will be considered first. For $\forall s$ with $\text{Re}(s)\geq 0$, the following inequalities hold:
\begin{align}
\Vert \mathcal{E}+\hat{\mathcal{F}}(s) \Vert &\leq \Vert \mathcal{E}\Vert+\Vert \hat{\mathcal{F}}(s) \Vert \\ \nonumber
                                                                        &\leq \Vert \mathcal{E}\Vert+\int_0^\infty |e^{-st}|\Vert \mathcal{F}(t)\Vert dt \\ \nonumber
                                                                        &\leq \mathcal{E}+\mathcal{F}_+. 
\end{align}
Let us denote the eigenvalues of $\mathcal{E}+\hat{\mathcal{F}}(s)$ by $\{\lambda_i(s)\}_{i=1}^N$.It is clear that $|\lambda_i(s)|\leq \Vert \mathcal{E}+\hat{\mathcal{F}}(s) \Vert $, for $\forall i$ (the 2-norm of a matrix is equal to its largest singular value). The eigenvalues of $sI_N-\mathcal{E}-\hat{\mathcal{F}}(s)$ are $\{s-\lambda_i(s)\}_{i=1}^N$. Clearly, for $\forall i$,
\begin{align}
|s-\lambda_i(s)|&\geq ||s|-|\lambda_i(s)|| \\ \nonumber
                       &> |\Vert \mathcal{E}\Vert+\mathcal{F}_+- \Vert \mathcal{E}+\hat{\mathcal{F}}(s) \Vert| \\ \nonumber
                       &\geq 0.
\end{align}
This implies that for $|s|>\Vert \mathcal{E}\Vert+\mathcal{F}_+$, $s-\lambda_i(s) \neq 0$, $\forall i$. Since $\text{Det}(sI_N-\mathcal{E}-\hat{\mathcal{F}}(s))=\prod_{i=1}^N (s-\lambda_i(s))$, we have $\text{Det}(sI_N-\mathcal{E}-\hat{\mathcal{F}}(s)) \neq 0$.

Then, the case where $|s|\leq \Vert \mathcal{E}\Vert+\mathcal{F}_+$ will be considered. Note that $\hat{\mathcal{F}}(0)=\int_0^\infty \mathcal{F}(t)dt$. The following inequalities hold for $s \neq 0$
\begin{align}
\Vert \hat{\mathcal{F}}(0)-\hat{\mathcal{F}}(s)\Vert &=\Vert \int_0^\infty (1-e^{-st})\mathcal{F}(t)dt \Vert \\ \nonumber
                                                                                    & \leq \int_0^\infty |1-e^{-st}|\Vert \mathcal{F}(t)\Vert dt \\ \nonumber
                                                                                    & \leq |s|\int_0^\infty t \Vert \mathcal{F}(t)\Vert dt \\ \nonumber
                                                                                    & < |s|\frac{2|\text{Re}(\lambda_1)|^N\mathcal{F}_+}{N(2\Vert \mathcal{E}\Vert+2\mathcal{F}_+)^N \mathcal{F}_+ },
\end{align}
where we have applied (C4) and the following inequality:
\begin{equation}
|1-e^{-z}|\leq |z|,\quad \text{Re}(z)\geq 0.
\end{equation}
Next, the spectral change between two complex $N\times N$ matrices $A$ and $B$ is introduced as
\begin{equation}
\text{SC}(A,B)=\max\limits_{1\leq j \leq N}\{\min\limits_{1\leq i \leq N} |\lambda_i^A-\lambda_j^B|\},
\end{equation}
where $\{\lambda_i^A\}_{i=1}^N$ and $\{\lambda_i^B\}_{i=1}^N$ denote the eigenvalues of $A$ and $B$, respectively. We also list the following notation for conciseness:
\begin{align*}
&\mathcal{F}_1(s)\triangleq sI_N-\mathcal{E}-\hat{\mathcal{F}}(s), \\ 
&\mathcal{F}_2(s)\triangleq \mathcal{E}+\hat{\mathcal{F}}(s). 
\end{align*}
According to (48) and the Bhatia-Friedland upper bound \cite{BHATIA19811}, we have for $s \neq 0$
\begin{align}
&\text{SC}(\mathcal{F}_2(0),\mathcal{F}_2(s)) \\ \nonumber 
&\leq N^{\frac{1}{N}}(2\hat{m})^{1-\frac{1}{N}}\Vert \hat{\mathcal{F}}(0)-\hat{\mathcal{F}}(s)\Vert^{\frac{1}{N}} \\ \nonumber
&< (2\hat{m})^{1-\frac{1}{N}}\frac{2^{\frac{1}{N}}|\text{Re}(\lambda_1)|}{2\Vert \mathcal{E}\Vert+2\mathcal{F}_+}|s|^{\frac{1}{N}},
\end{align}
where $\hat{m}=\max\{\mathcal{F}_2(0),\mathcal{F}_2(s)\}$.

Let $\lambda_1(s)$ be the eigenvalue of $\mathcal{F}_2(s)$ with the largest real part, which makes $s-\lambda_1(s)$ the eigenvalue of $\mathcal{F}_1(s)$ with the smallest real part. The definition of spectral change indicates the following ($s \neq 0$):
\begin{align}
&\text{Re}(s-\lambda_1(s)) \\ \nonumber 
&\geq \text{Re}(s-\lambda_1)-\text{SC}(\mathcal{F}_1(0),\mathcal{F}_1(s)) \\ \nonumber
&= \text{Re}(s-\lambda_1)-\text{SC}(\mathcal{F}_2(0),\mathcal{F}_2(s)) \\ \nonumber
&>|\text{Re}(\lambda_1)|-(2\hat{m})^{1-\frac{1}{N}}\frac{2^{\frac{1}{N}}|\text{Re}(\lambda_1)|}{2\Vert \mathcal{E}\Vert+2\mathcal{F}_+}|s|^{\frac{1}{N}} \\ \nonumber
&\geq |\text{Re}(\lambda_1)|-2^{1-\frac{1}{N}}(\Vert \mathcal{E}\Vert+\mathcal{F}_+)\frac{2^{\frac{1}{N}}|\text{Re}(\lambda_1)|}{2\Vert \mathcal{E}\Vert+2\mathcal{F}_+} \\ \nonumber
&=0.
\end{align}
Therefore, for $s \neq 0$, all eigenvalues of $sI_N-\mathcal{E}-\hat{\mathcal{F}}(s)$ must have real parts strictly greater than 0, which makes themselves nonzero. Moreover, since $\mathcal{E}+\hat{\mathcal{F}}(0)$ is Hurwitz, $sI_N-\mathcal{E}-\hat{\mathcal{F}}(s)$ has nonzero eigenvalues for all $s$ with $|s|\leq \Vert \mathcal{E}\Vert+\mathcal{F}_+$, and $\text{Re}(s)\geq 0$. 

At this point, we have shown that $\text{Det}(sI_N-\mathcal{E}-\hat{\mathcal{F}}(s))\neq 0$ for all $s$ with $\text{Re}(s)\geq 0$. Therefore, for all $s$ on the right half complex plane and the imaginary axis, $sI_N-\mathcal{E}-\hat{\mathcal{F}}(s)$ is invertible and
\begin{equation}
\hat{y}(s)=(sI_N-\mathcal{E}-\hat{\mathcal{F}}(s))^{-1}y(0).
\end{equation}
Since all poles of $\hat{y}(s)$ must lie on the open left half complex plane, we invoke the final value theorem \cite{4213171} and yield that 
\begin{equation}
\lim\limits_{t\to \infty}y(t)=\lim\limits_{s\to 0}s\hat{y}(s)=0,
\end{equation}
which concludes the proof.
\end{pf}

{\bfseries Remark 2.} \textit{Since we have assumed the integrability of the memory kernel matrix, its norm can be viewed as a distribution density over the real line after normalization. Consider a sequence of such distribution with increasing mass deployed at the origin. Then, in the limit of Dirac} $\delta$\textit{, we reach an ordinary linear system with Hurwitz system matrix, which is asymptotically stable. Then, it can be reasonably expected that this stability might be retained if the distribution does not deviate too far from} $\delta$. \textit{The amount of deviation is measured in this work by the left-hand side of (44), which is the expectation of the distribution concerned. It should be kept below a certain threshold given that the impulsive distribution has expectation 0. Moreover, this quantity is an upper bound of the Earth Mover distance between the concerned distribution and Dirac} $\delta$ \textit{distribution}.

We are now in the position to present the main result.

\begin{thm}
Suppose that $(\Omega_1,V_1,\Gamma_1(t))=(\Omega_2,V_2,\Gamma_2(t))$. Denote by $\Gamma_n(t)$ the top left $n\times n$ block of $\Gamma_i(t)$, $i=1,2$. Let $\Gamma_n \triangleq \int_0^\infty \Gamma_n(t)dt$,  $\Gamma_{n+} \triangleq \int_0^\infty \Vert\Gamma_n(t)\Vert dt$ and $p_{\Gamma_n}(t)\triangleq  \Vert\Gamma_n(t)\Vert /\Gamma_{n+}$. Synchronization in quantum expectation sense can be engineered if there exists a real number $a>\Vert J_n \Omega_1 \Vert$, s.t.
\begin{align}
&\int_0^\infty t p_{\Gamma_n}(t)dt < \\ \nonumber
&\frac{|\text{Re}(\lambda_{1}(J_n \Omega_1))-a|^{2n}\sigma_{\min}(\Gamma_n)}{n(2\Vert J_n\Omega_1\Vert+2a\frac{\Gamma_{n+}}{\sigma_{\min}(\Gamma_n)})^{2n}(2a\frac{\Gamma_{n+}}{\sigma_{\min}(\Gamma_n)})\sigma_{\max}(\Gamma_n)},
\end{align}
where $\lambda_{1}(J_n \Omega_1)$ stands for the eigenvalue of $J_n \Omega_1$ with the largest real part, and $\sigma_{\min}(\Gamma_n)$, $\sigma_{\max}(\Gamma_n)$ represent the smallest and largest singular value of $\Gamma_n$, respectively. 
\end{thm}

\begin{pf}
We employ a constructive approach to prove the conclusion. System parameters are explicitly designed to demonstrate that Definition 1 is satisfied. 

Suppose that $a$ is the positive real number that guarantees (55). We construct the following matrix:
\begin{equation}
K=\sqrt{a}\begin{pmatrix} K_1 \\ K_2 \end{pmatrix},
\end{equation}
where $K_1=I_M \otimes (1\quad \text{i})$ and $K_2=0_{(M-n)\times 2n}$. Let $\Gamma=\int_0^\infty \Gamma_i(t)dt$, $i=1,2$. The remaining parameters in the augmented system are constructed as:
\begin{align}
V_{12}&=V_1-\Gamma^{-\frac{1}{2}}K, \\ \nonumber
V_{21}&=V_{12}, \\ \nonumber
\Omega_{12}&=0 .
\end{align}
Given the conditions of Theorem 2 and the parameters above designed, it can be checked that (29) and (30) are satisfied, which leads to (C1) due to their sufficiency. We then  proceed to show that (C2) is satisfied as well.

It can be verified that system matrices in (40) governing the evolution of synchronization error are:
\begin{align}
E&=2J_n\Omega_1, \\ \nonumber
F(t)&=2aJ_n\text{Im}[(V_1-V_{12})^\dagger \Gamma_1(t)(V_1-V_{12})]\\ \nonumber
      &=2aJ_n\text{Im}(K^\dagger \Gamma^{-\frac{1}{2}}\Gamma_1(t) \Gamma^{-\frac{1}{2}} K).
\end{align}
Let $F \triangleq \int_0^\infty F(t)dt$. The chosen parameters in (57) yield:
\begin{align}
F&=2aJ_n\text{Im}(K^\dagger  K) \\ \nonumber
  &=2aJ_n\text{Im}\Bigg{[}I_n \otimes \begin{pmatrix} 1 & \text{i} \\ -\text{i} & 1 \end{pmatrix}\Bigg{]}\\ \nonumber
  &=2a J_n^2 \\ \nonumber
  &=-2a I_{2n}.
\end{align}
Therefore,
\begin{equation}
E+F=2J_n\Omega_1-2a I_{2n}
\end{equation}
is Hurwitz since $a>\Vert J_n \Omega_1 \Vert$. The assumption in Lemma 1 is thus fulfilled.

Let us write $\Gamma_1(t)=\text{diag}(\gamma_1(t),...,\gamma_M(t))$ and $\Gamma=\text{diag}(\gamma_1,...,\gamma_M)$. Similar to (59), we also derive the following expression:
\begin{equation}
F(t)=\text{diag}(\frac{\gamma_1(t)}{\gamma_1}I_2,...,\frac{\gamma_n(t)}{\gamma_n}I_2).
\end{equation}
Given this explicit expression, we have the following relationships:
\begin{align}
\Vert F(t) \Vert&=2a\max\limits_{1\leq i \leq n}\{\frac{\gamma_i(t)}{\gamma_i}\}=\frac{\gamma_{i_0}(t)}{\gamma_{i_0}} \\ \nonumber
\Vert \Gamma_n(t) \Vert&=\max\limits_{1\leq j \leq n}\{\gamma_j(t)\}= \gamma_{j_0}(t).
\end{align}
It then holds that
\begin{align}
\Vert F(t) \Vert \sigma_{\min}(\Gamma_n)&=2a\frac{\gamma_{i_0}(t)}{\gamma_{i_0}}\sigma_{\min}(\Gamma_n) \\ \nonumber
                                                                 &\leq 2a \gamma_{i_0}(t) \\ \nonumber
                                                                 &\leq 2a \Vert \Gamma_n(t)\Vert,
\end{align}
and that 
\begin{align}
2a \Vert \Gamma_n(t)\Vert&=2a\gamma_{j_0}(t) \\ \nonumber
                                         &\leq 2a \frac{\gamma_{j_0}(t)}{\gamma_{j_0}}\sigma_{\max}(\Gamma_n) \\ \nonumber
                                         &\leq \Vert F(t) \Vert \sigma_{\max}(\Gamma_n).
\end{align}
Let $F_+ \triangleq \int_0^\infty \Vert F(t) \Vert dt$ and $p_F(t) \triangleq \Vert F(t) \Vert / F_+$. By virtue of (63) and (64), it becomes clear that
\begin{align}
&\quad \int_0^\infty t p_F(t)dt \\ \nonumber
&\leq \int_0^\infty t p_{\Gamma_n}(t)dt \frac{\sigma_{\max}(\Gamma_n)}{\sigma_{\min}(\Gamma_n)}\\ \nonumber
&< \frac{|\text{Re}(\lambda_{1}(J_n \Omega_1))-a|^{2n}}{n(2\Vert J_n\Omega_1\Vert+2a\frac{\Gamma_{n+}}{\sigma_{\min}(\Gamma_n)})^{2n}(2a\frac{\Gamma_{n+}}{\sigma_{\min}(\Gamma_n)})}\\ \nonumber
&\leq \frac{|\text{Re}(\lambda_{1}(J_n \Omega_1))-a|^{2n}}{n(2\Vert J_n\Omega_1\Vert+F_+)^{2n}F_+}\\ \nonumber
&= \frac{2|\text{Re}(\lambda_{1}(E+F))|^{2n}}{2n(2\Vert E\Vert+2F_+)^{2n}F_+}.
\end{align}
Eq. (65) implies that both (C3) and (C4) are satisfied, which validates the applicability of Lemma 1. We thus finally obtain $\lim\limits_{t\to \infty}e(t)=0$. The proof is  completed.
\end{pf}
{\bfseries Remark 3.} \textit{Similar to Remark 2, the left-hand side of ~(55) can be viewed as a measure of non-Markovianity for the system. The larger the expectation is, the more evenly the memory kernel is smeared over the timeline, and the more significant memory effect is. We have shown that synchronization between two subsystems is possible if the degree of non-Markovianity is contained below a certain level. Also, it is noticable that direct Hamiltonian coupling is not needed in our construction. Synchronization can be synthesized only through mutual coupling to the environment. This is in some way analogous to Huygens' pendulum.}

In Theorem 2, (56) and (57) present specific parameters to ensure synchronization. It can be inferred from (55) that, if a set of parameters work for two given subsystems, they will also work for any subsystem pairs with less significant non-Makovianity. We outline this conclusion in the following corollary and omit its proof since the proof is straightforward.

{\bfseries Corollary 3.} \textit{Suppose that engineered parameters in (56) and (57) lead to synchronization in quantum expectation sense for the two subsystems as described in Theorem 2.} \textit{Let the memory kernels of the two subsystems be replaced as} $\Gamma'_j(t)$, $j=1,2$ \textit{with} $\Gamma'_1(t)=\Gamma'_2(t)$. \textit{Denote} $\Gamma'_n(t)$ \textit{as the top left} $n\times n$ \textit{block of} $\Gamma'_j(t)$, $j=1,2$. \textit{Let} $\Gamma'_n \triangleq \int_0^\infty \Gamma'_n(t)dt$, $\Gamma'_{n+} \triangleq \int_0^\infty \Vert\Gamma'_n(t)\Vert dt$ and $p_{\Gamma'_n}(t)\triangleq  \Vert\Gamma'_n(t)\Vert /\Gamma'_{n+}$. \textit{Then, parameters in (56) and (57) still lead to synchronization if} $\Gamma'_n=\Gamma_n$ \textit{and}
\begin{equation}
\int_0^\infty t p_{\Gamma'_n}(t)dt \leq \int_0^\infty t p_{\Gamma_n}(t)dt.
\end{equation}

Here we also provide a few remarks on the relation between our work and coherent quantum observers \cite{6981590,7963093,MIAO2016264}. Although both topics involve some form of tracking between subsystems, coherent observers aim at extracting information from and performing estimation on the plant system, while our work focuses on engineering the subsystems to synchronization. In this sense, the former views the two subsystems as different in terms of functionality while the latter treats them on an equal footing. If the observer is allowed to change inherent plant dynamics, our work can also be applied to the design of coherent observer.

\section{Numerical Example}
In this section, we give an example to illustrate our theoretical results. The original parameters of the two subsystems $(\Omega_i,V_i,\Gamma_i(t))$, $i=1,2$, are :
\begin{align}
&\Omega_1=\Omega_2=\begin{pmatrix}0 & 0.1 \\ 0.1 & 0 \end{pmatrix},\\ \nonumber
&V_1=V_2=\begin{pmatrix}0.2 &\ \ -0.1\text{i}  \end{pmatrix},\\ \nonumber
&\Gamma_1(t)=\Gamma_2(t)=\beta e^{-\beta t},
\end{align}
where we set $\beta=9$. Such an exponential kernel function appears when input noise admits a Lorentz-type spectrum \cite{PhysRevA.87.032117}.

Let $K=\sqrt{0.4}(1 \quad \text{i})$. The remaining parameters are designed as:
\begin{align}
V_{12}&=V_1-K \\ \nonumber
            &=\begin{pmatrix}0.2-\sqrt{0.4} &\ \ -0.1\text{i}-\sqrt{0.4}\text{i}  \end{pmatrix},
\end{align}
and
\begin{equation}
V_{21}=V_{12},\quad \Omega_{12}=0.
\end{equation}
Moreover, there exists $a=0.4>\Vert J\Omega_1\Vert=0.1$, s.t 
\begin{align}
\int_0^\infty t p_{\Gamma_n}(t)dt &= \frac{1}{9} \\ \nonumber
                                                       &< \frac{|0.1-0.4|^2}{(0.2+0.8)^2 \times 0.8}\\ \nonumber
                                                       &= \frac{9}{80}.
\end{align}
The conditions in Theorem 2 are all met. We choose the following three initial states for the augmented system
\begin{align}
\rho_{\text{sys},1}&=|\alpha_1\rangle \langle \alpha_1| \otimes |\alpha_2\rangle \langle \alpha_2| \\ \nonumber
\rho_{\text{sys},2}&=|\alpha_2\rangle \langle \alpha_2| \otimes |\alpha_3\rangle \langle \alpha_3| \\ \nonumber
\rho_{\text{sys},3}&=|\alpha_1\rangle \langle \alpha_1| \otimes |\alpha_3\rangle \langle \alpha_3|,
\end{align}
where $\alpha_1=1$, $\alpha_2=0$, $\alpha_3=\text{i}$, and $|\alpha_k\rangle$ represents coherent state with eigenvalue $\alpha_k$, $k=1,2,3$. The evolution of synchronization error $|e|=\sqrt{e_1^2+e_2^2}$ is plotted in Fig.1. It can be seen that the error vanishes aymptotically for all three initial states.
\begin{figure}
\begin{center}
\includegraphics[height=7cm]{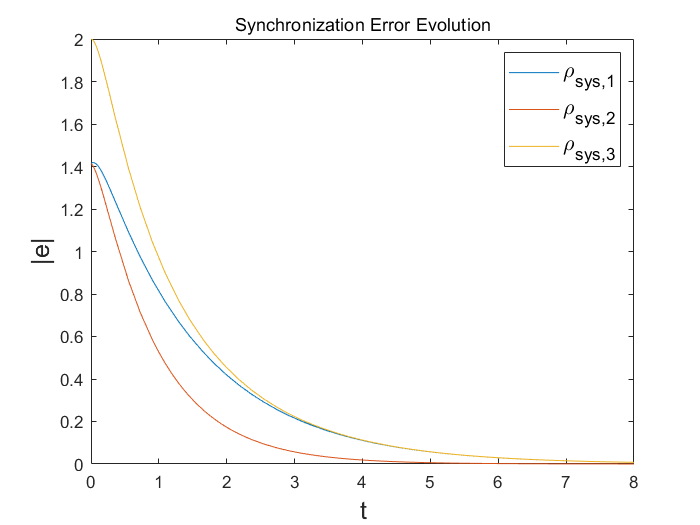}    
\caption{The evolution of synchronization error is asympotically stable.}  
\label{fig1}                                 
\end{center}                                 
\end{figure}

\section{Conclusion and Discussion}
For non-Markovian linear quantum systems, we have established a set of sufficient conditions under which synchronization in quantum expectation sense can be achieved. Specific parameters of the augmented system are explicitly constructed to engineer synchronization. The results may be useful in quantum information processing tasks involving non-Markovian quantum devices.

From a theoretical perspective, non-Markovian quantum dynamics may harbor features regarding \textit{nonclassicality} which are not present in their Markovian counterparts. For example, it has been recently shown that certain quantum processes with memory show nonclassical statistics regardless of the measurement scheme, as opposed to memoryless dynamics \cite{PhysRevX.10.041049}. To further explore the possibilities of nonclassicality in the non-Markovian setting, it might be interesting to study the nonclassical correlations between synchronized systems.

\begin{ack}                               
This work was supported by the Open Subject of Beijing Intelligent Logistics System Collaborative Innovation Center under Grant BILSCIC-2019KF-13, in part by the State Key Laboratory of Synthetical Automation for Process Industries. It is also partially supported by the Australian Research Council’s Discovery Projects funding scheme under Project DP190101566.  
\end{ack}






\end{document}